\documentclass[]{pasj01}
\draft

\Received{}
\Accepted{}
 
\begin{document}

\title{Photometric and Spectroscopic Properties of the Eclipsing System V864 Monocerotis}

\author{Jang-Ho \textsc{Park}\altaffilmark{1}%
}
\altaffiltext{1}{Korea Astronomy and Space Science Institute, Daejeon 34055, Republic of Korea}
\altaffiltext{2}{Astronomical Institute, Faculty of Mathematics and Physics, Charles University in Prague, 180 00 Praha 8, V Hole\v sovi\v ck\'ach 2, Czech Republic}
\email{pooh107162@kasi.re.kr}

\author{Jae Woo \textsc{Lee}\altaffilmark{1,2}}
\author{Kyeongsoo \textsc{Hong}\altaffilmark{1}}

{}

\maketitle

\begin{abstract}
We present the orbital period variability and evolutionary status of the W UMa-type binary system V864 Mon from accurately measured fundamental parameters.
New $BV$ photometric observations of this system were performed in 2019 January and 2022 January, and
the first high-resolution spectroscopic observations were carried out on three nights between 2019 January and March.
A total of 29 times of minimum light were collected to examine the behavior of the orbital period.
Our analysis of these timings indicates a continuous period increase at a rate of $+$2.62$\times$10$^{-7}$ d yr$^{-1}$ over the past 20 years,
which can be interpreted as a mass transfer from the less massive primary to the secondary component with a rate of 1.22$\times$10$^{-7}$ M$_\odot$ yr$^{-1}$.
We measured the radial velocities (RVs) for both components, and determined the effective temperature and projected rotational velocity of the more massive secondary star to be $T_{\rm eff,2}$ = 5450 $\pm$ 94 K and $v_2 \sin i$ = 192 $\pm$ 40 km s$^{-1}$, respectively,
from the comparison of the observed spectrum at the primary minimum and the theoretical models.
The individual masses and radii of both components were determined from a simultaneous analysis of the light and RV curves,
which are $M_1$ = 0.34 $\pm$ 0.02 M$_\odot$, $R_1$ = 0.69 $\pm$ 0.01 R$_\odot$, and $M_2$ = 1.06 $\pm$ 0.04 M$_\odot$, $R_2$ = 1.16 $\pm$ 0.02 R$_\odot$, respectively.
Our results indicate that V864 Mon is a W-subtype of W UMa stars with time-varying spot activity.
The positions in the mass-luminosity and mass-radius diagrams indicate that the secondary star belongs to the main sequence region, while the hotter primary is located beyond the terminal-age main sequence.
\end{abstract}

\section{Introduction}

W UMa-type binaries generally consist of two dwarfs, and because the distance between both components are very close they usually have a short orbital period of 0.3 to 0.8 days.
They are in contact with each other and are surrounded by a common envelope.
Also, because the surface temperatures are nearly equal, the eclipse depths of both minima appear similar.
Binnendijk (1970) classified the W UMa stars into two subtypes based on which component is eclipsed at the deeper primary minimum.
At the primary eclipse, when the less massive and cooler component transit the more massive and hotter component, it is classified as an A-subtype, and
conversely, if the less massive but hotter component is occulted by its companion, it is classified as a W-subtype.
The systems belonging to A-subtype are mainly known as early-type (A-F), and W-subtype as late-type (G-K).
The W UMa-type binaries are very interesting objects whose structure, formation and evolution are not yet well understood (Webbink 2003; Eggleton 2012; St\c {e}pie\'{n} \& Gazeas 2012).
The fundamental parameters of the contact systems (e.g., mass and radius) are important factors in understanding not only their characteristics but also their current evolutionary status.
By analyzing precise photometric and spectroscopic data together, more accurate fundamental parameters can be obtained,
which in turn can contribute to constructing more advanced solutions for understanding their complex characteristics.

This paper is concerned with V864 Mon (HD 296761, Gaia DR3 3059296928547426688, TYC 4824-153-1), which was identified as a stellar object with a spectral type of G0 and a magnitude of $B$ = 10.6 mag (Nesterov et al. 1995).
Wils \& Dvorak (2003) classified this target as a W UMa-type binary with an orbital period of 0.35843 day.
Szczygie{\l} et al. (2008) calculated their distance to be 151.6 pc from a distance modulus.
Avvakumova et al. (2013) classified this system as a W-subtype of W UMa star.
From the distance-independent colours $G_{\rm BP} - G$ and $G - G_{\rm RP}$, its effective temperature was estimated to be 5860 K (Gaia Collaboration et al. 2018).

Despite its nearly 30-year observational history V864 Mon is a neglected binary system whose photometric and spectroscopic properties remain largely unknown.
Therefore, the main purpose of this paper is to examine the orbital period variability of this system and
to investigate their evolutionary status with the accurate fundamental parameters measured through a simultaneous analysis of our new $BV$ light curves and high-resolution spectroscopic data.
The remainder of this paper is organized as follows:
Section 2 describes our observations and data reductions.
Section 3 reports the results of the orbital period analysis.
Section 4 explains the measurements of the RVs and atmospheric parameters from the spectroscopic data.
The fundamental parameters of each component are determined from the binary modeling in Section 5.
Finally, in Section 6 our results are summarized and discussed.

\section{Observations and Data Reductions}

New CCD photometric observations of V864 Mon were obtained during two observing seasons, 2019 January and 2022 January,
using a FLI 4K CCD camera and a $BV$ filter set attached to the 61 cm reflector at the Sobaeksan Optical Astronomy Observatory (SOAO) in Korea.
The CCD chip used is 9 $\mu$m in size and has 4096 $\times$ 4096 pixels, and the field of view (FoV) is 15$\arcmin$.2 $\times$ 15$\arcmin$.2.
All CCD frames were corrected for flat, bias, and dark images with the IRAF/CCDRED package, and then aperture photometry was performed on most of the stars in our observing field with the IRAF/PHOT package.
Among the stars with a color index similar to V864 Mon, TYC 4823-3121-1 and TYC 4823-3097-1,
which were stable without variations in brightness during our observing period, were selected as comparison (C) and check (K) stars, respectively.
The standard deviations of the (K$-$C) differences (1$\sigma$) are about $\pm$ 0.006 mag in both bands.
As a result, we obtained a total of 1051 individual points (521 in $B$, 530 in $V$) during the two observing seasons.

Spectroscopic observations were made on three nights between 2019 January and March using the 1.8 m reflector and
the Bohyunsan Optical Echelle Spectrograph (BOES) at the Bohyunsan Optical Astronomy Observatory (BOAO) in Korea.
Among five optical fibers of BOES that cover the wavelength region of 3600 to 10,200 $\rm \AA$ (Kim et al. 2007),
we used the largest fiber (300 $\mu$m) with a resolution of $R$ = 30,000 to maximize the signal-to-noise (S/N) ratios of the spectra.
The observations were carried out with an exposure of 600 s considering the brightness and orbital period of the object, and a total of 26 high-resolution spectra were obtained.
These were preprocessed through the IRAF/CCDRED package, and one-dimensional spectra were extracted with the IRAF/ECHELLE package.
The S/N ratios at $5000-6000$ $\rm \AA$ were approximately 40.

\section{Orbital Period Analysis}

From our photometric observations, seven weighted times of minimum light and their errors were determined using the method of Kwee \& van Woerden (1956).
Including ours, a total of 29 timings were collected from the database published by Kreiner et al. (2001) and more recent literature, and they are listed in Table 1.
Because some of the timings had been published without errors, they were assigned the standard deviations of the full $O$--$C$ residuals.

In order to figure out the behavior of the orbital period of V864 Mon, an eclipse timing $O$--$C$ diagram was constructed using Kreiner et al.'s ephemeris:
\begin{equation}
 C_1 = \mbox{HJD}~ 2,453,511.4873 + 0.3584255E.
\end{equation}
This is plotted in the upper panel of Figure 1, and the $O$--$C_{1}$ residuals are listed in the fourth column of Table 1,
where the timings are marked as a down-pointing triangle (visual) and squares (CCD) according to their observational methods.
The diagram shows that the orbital period of this system has increased continuously in the form of an upward parabola over about 20 years.
Therefore, we introduced a parabolic least-squares fit for all times of minimum light and obtained the following ephemeris:
\begin{equation}
 C_2 = \mbox{HJD}~ 2,453,511.4872(42) + 0.3584246(9)E + 1.29(48) \times 10^{-10}E^2.
\end{equation}
The result is represented as a continuous curve in the upper panel of Figure 1.
The $O$--$C_{2}$ residuals from the quadratic ephemeris are plotted in the lower panel of Figure 1, and they are listed in the fifth column of Table 1.

As shown in Figure 1, all of the eclipse minima of V864 Mon fit well with the upward parabolic ephemeris.
The secular variation represents a continuous period increase at a rate of $+$2.62$\times$10$^{-7}$ d yr$^{-1}$,
corresponding to a fractional period change of $+$7.18$\times$10$^{-10}$.
The period variation could be explained by assuming conservative mass transfer between the V864 Mon components.
The transfer rate from the less massive to the more massive component is computed as 1.22$\times$10$^{-7}$ M$_\odot$ yr$^{-1}$
from the relation between the period increase rate and the masses of both components (see Section 5).

\section{Spectral Analysis}

To measure the radial velocities (RVs) of V864 Mon, the following procedure was performed.
First, trail spectra were made from all the observed data, and we searched the absorption regions in which the orbital motions of both components were well shown.
We found double sinusoids in the Mg I $\lambda$5183.61, H$_{\alpha}$ $\lambda$6562.81, and Ca II $\lambda$8662 regions, and
identified the measurable absorption lines in each region.
Among them, the absorption line of Mg I $\lambda$5183.61 more clearly defined the orbital motion of the two components, so we chose that spectral region for the RV measurement.
Then, using the IRAF $splot$ task, the RVs of each component were measured about five times with double Gaussian functions.
The results are listed in Table 2, where the RVs and their errors are given as the averages and standard deviations ($\sigma$) of the measured values, respectively.

Our binary star model, presented in Section 5, indicates that the less massive component was completely occulted by the heavier and larger component at the primary minimum.
Therefore, to determine the effective temperature ($T_{\rm eff}$) and projected rotational velocity ($v \sin i$) of the more massive component, we used a spectrum observed at the orbital phase of 0.01.
As mentioned in the Introduction, V864 Mon was classified as an G-type star.
Therefore, we adopted four regions (Ca I $\lambda$4226, Fe I $\lambda$4325, Fe I $\lambda$4383, and Mg II $\lambda$4481) suitable as the temperature indicators of G-type dwarfs,
from the \textit{Digital Spectral Classification Atlas} of R. O. Gray\footnote{More information is available on the website: https://ned.ipac.caltech.edu/level5/Gray/frames.html}.
Then, to find the best atmospheric parameters for the observed spectrum,
a total of 49,966 synthetic spectra in ranges of ${T_{\rm eff} = 4000-7000}$ K and $v \sin i = 85-250$ km s$^{-1}$ were generated from the ATLAS-9 atmosphere models of Kurucz (1993).
In this procedure, the surface gravity (log $g$) was adopted from our binary parameters, and
the solar metallicity (Fe/H) and microturbulent velocity were assumed to be 0.0 and 2.0 km s$^{-1}$, respectively.

Finally, we used the $\chi^2$ fitting method between the observed spectrum and the synthetic spectra to find the optimal values for both atmospheric parameters.
This process is the same as that applied by Hong et al. (2021) and Park et al. (2020).
The optimal parameters for the observed spectrum were determined to be $T_{\rm eff,2}$ = 5450 $\pm$ 94 K and $v_2 \sin i$ = 192 $\pm$ 40 km s$^{-1}$, respectively, and
the result is presented in Figure 2.

\section{Binary Modeling}

To obtain the binary parameters of V864 Mon, our $BV$ light and double-lined RV curves were analyzed together.
The binary modeling was performed using mode 3 for overcontact binaries among operation modes provided by the Wilson-Devinney synthesis code (Wilson \& Devinney 1971; van Hamme \& Wilson 2003; hereafter W-D).
Here, the effective temperature of the secondary star ($T_{\rm 2}$) was fixed at 5450 K as determined by the spectral analysis.
The gravity-darkening exponents ($g$) and the bolometric albedos ($A$) of both components were adopted as values of 0.32 and 0.5 (Lucy 1967; Rucinski 1969),
respectively, considering their temperatures presented in Table 3.
The linear bolometric ($X$, $Y$) and monochromatic ($x$, $y$) limb-darkening coefficients were interpolated from the values of van Hamme (1993).

As shown in Figure 3, V864 Mon shows light variations even over a short period of the observations.
In order to examine the variations, the lights in 2019 and 2022 were compared for each phase (0.00, 0.25, 0.50, and 0.75).
Phases 0.25 and 0.75 have observational scatters but similar lights on average, and the secondary minimum (0.50) was almost identical to each other.
On the other hand, the primary minimum (0.00) in both bands showed a magnitude difference of about 0.025.
We first simultaneously analyzed the $BV$ light and RV curves of 2019 obtained during the same period.
In this process, the adjustment parameters considered were the orbital inclination $i$, the temperature of the primary star $T_1$, the dimensionless surface potential $\Omega_1$, the epoch $T_0$, the orbital period $P$, and
the monochromatic luminosity of the primary star $L_1$ for photometric parameters and the semimajor axis $a$, the system velocity $\gamma$, and the mass ratio $q$ for spectroscopic parameters.
However, the synthetic curves did not fit well the observed ones at the primary minimum.
This discordance at the eclipse can be caused by time-varying spot activity (e.g., Guo et al. 2018 for V474 Cam; Zhu et al. 2019 for V1005 Her).
Therefore, the spot parameters that can reasonably explain this discrepancy were included in the photometric parameters.
The results are indicated by the circles and solid lines in Figure 3, and it can be seen that our observations and synthetic model curves are in good agreement.
Figure 4 shows the RV curves of V864 Mon with the model fits.
Binary and spot parameters are listed in Table 3 and columns (2)--(3) of Table 4, respectively.
To examine the time-varying spot activity, the light curves of 2022 were analyzed by adjusting only $T_0$, $L_1$, and the spot parameters.
The results are indicated by plus symbols and dashed lines in Figure 3, and listed in columns (4)--(5) of Table 4.
Figure 5 shows the three-dimensional Roche geometry and spot activity of V864 Mon at four different phases.
All images were implemented by applying our binary solutions to the Binary Maker 3.0 program (Bradstreet \& Steelman 2002).

The fundamental parameters were determined by applying our light and RV parameters to the JKTABSDIM code (Southworth et al. 2005).
Here, the luminosities ($L$) and bolometric magnitudes ($M_{\rm bol}$) were calculated with the solar values of $T_{\rm eff}$$_\odot$ = 5777 K and $M_{\rm bol}$$_\odot$ = +4.75 provided by Zombeck (1990).
The absolute visual magnitudes ($M_{V}$) were derived using the bolometric corrections (BCs) of Girardi et al. (2002), suitable for the effective temperatures of each component.
The results are listed in the bottom of Table 3.

\section{Summary and Discussion}

In this paper, the new photometric and spectroscopic observations of the W UMa-type contact binary system V864 Mon were presented.
We analyzed these two kinds of observational data together, and the results can be summarized as follows:

\begin{enumerate}

\item The orbital period of V864 Mon continuously increases in the form of an upward parabola with a rate of $+$2.62$\times$10$^{-7}$ d yr$^{-1}$,
which can be interpreted as a mass transfer from the less massive component to the more massive component with a rate of 1.22$\times$10$^{-7}$ M$_\odot$ yr$^{-1}$.

\item We measured the RVs in the Mg I $\lambda$5183.61 region, where the absorption lines of the two components are well separated.
The spectroscopic mass ratio of this system was calculated to be 3.122 $\pm$ 0.080 from their RV semi-amplitudes ($K_1$, $K_2$).

\item The effective temperature and projected rotational velocity of the more massive secondary component were derived using the observed spectrum at the primary minimum where only the more massive secondary star was visible.
We found the optimal values between the observed and synthetic spectra utilizing the $\chi^2$ fitting method.
The best atmospheric parameters were determined to be $T_{\rm eff,2}$ = 5450 $\pm$ 94 K and $v_2 \sin i$ = 192 $\pm$ 40 km s$^{-1}$.

\item Through a simultaneous analysis of the light and RV curves, we obtained the binary parameters of V864 Mon to be $a$ = 2.37 R$_\odot$, $i$ = 80.7 deg, a temperature difference of $\Delta T$ = 17 K, and a fill-out factor of $f$ = 13 \%.
The fundamental parameters of this system are as follows:
$M_1$ = 0.34 $\pm$ 0.02 M$_\odot$, $R_1$ = 0.69 $\pm$ 0.01 R$_\odot$ and $M_2$ = 1.06 $\pm$ 0.04 M$_\odot$, $R_2$ = 1.16 $\pm$ 0.02 R$_\odot$.

\item The seasonal light curves show the change of eclipse depth at the primary minimum.
This can be satisfactorily explained as the phenomenon of spot activity on the secondary star,
whose radius and temperature ($T$$\rm _{spot}$/$T$$\rm _{local}$) varies drastically with time.

\end{enumerate}

To investigate the evolutionary status of V864 Mon, both components are plotted in the mass-luminosity ($M-L$, left) and mass-radius ($M-R$, right) diagrams in Figure 6.
In this figure, the zero-age main sequence (ZAMS) and terminal-age main sequence (TAMS) lines were obtained from Girardi et al. (2000) for the solar chemical composition.
In addition, other 195 W-subtype of W UMa stars (Latkovi\'c et al. 2021) were plotted together for comparison.
V864 Mon can be seen to have physical properties similar to most W-subtype contact binaries.
The more massive secondary belongs to the main sequence region, and the primary component is located beyond the TAMS,
which is relatively more evolved, and is brighter and larger compared to its own mass.
The main cause of this may be the result of energy transfer from the more massive component toward the less massive component during their evolutionary process (Lucy 1968; Li et al. 2008; Zhang et al. 2020).
W UMa-type contact binaries undergo a periodic evolution process called thermal relaxation oscillation (TRO, Flannery 1976; Lucy 1976; Robertson \& Eggleton 1977).
In a state of evolution from contact to semi-detached states, mass transfers from the less massive component to the more massive component,
opposite to energy transfer, and an orbital period increase occurs.
Based on TRO, our results suggest that V864 Mon currently is in an expanding state before broken contact phase.

\begin{ack}
This paper is based on observations obtained at SOAO and BOAO, which are operated by the Korea Astronomy and Space Science Institute (KASI).
We would like to thank the staffs of both sites for assistance during our observations.
We also thanks Prof. Chun-Hwey Kim for providing us the times of minimum light for V864 Mon.
We have used the Simbad database maintained at CDS, Strasbourg, France.
This work was supported by KASI grant 2023-1-832-03.
The work by K. Hong was supported by the grant Nos. 2020R1A4A2002885 and 2022R1I1A1A01053320 from the National Research Foundation (NRF) of Korea.
\end{ack}


\clearpage
\begin{figure}
\includegraphics[width=1\columnwidth]{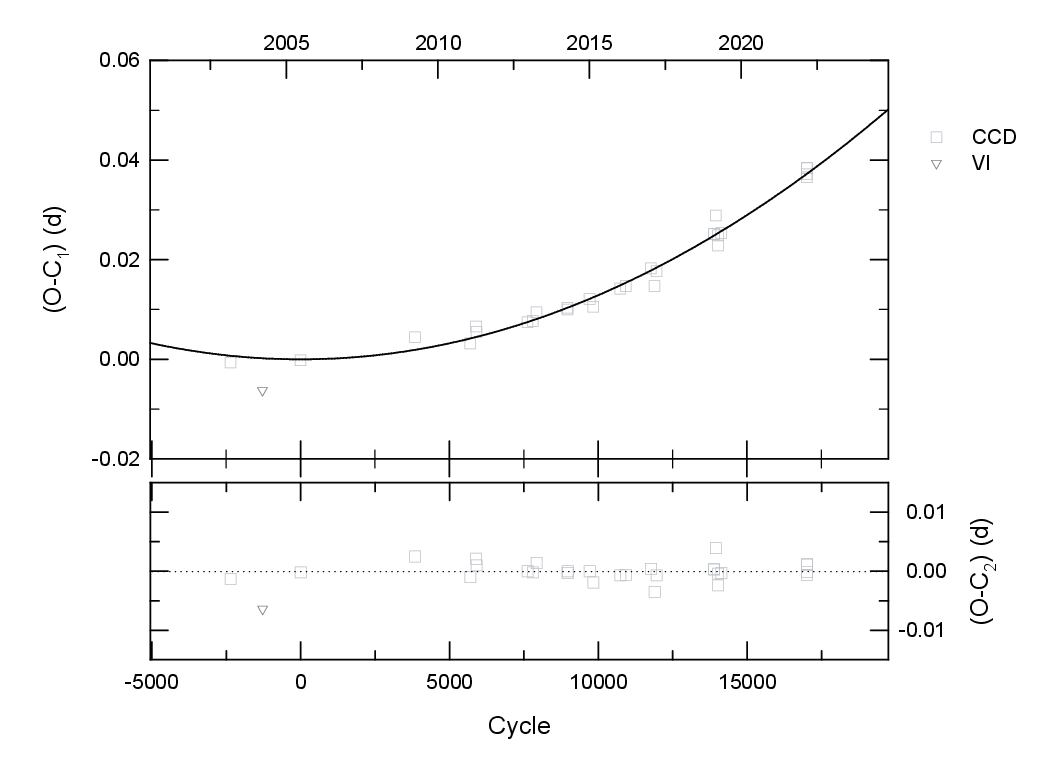}
\caption{Eclipse timing $O$--$C$ diagram of V864 Mon.
In the upper panel, constructed with the Kreiner et al.'s ephemeris, the continuous curve represents the quadratic term of equation (2).
The lower panel shows the residuals from the quadratic ephemeris.}
\label{Fig1}
\end{figure}

\clearpage
\begin{figure}
\includegraphics[]{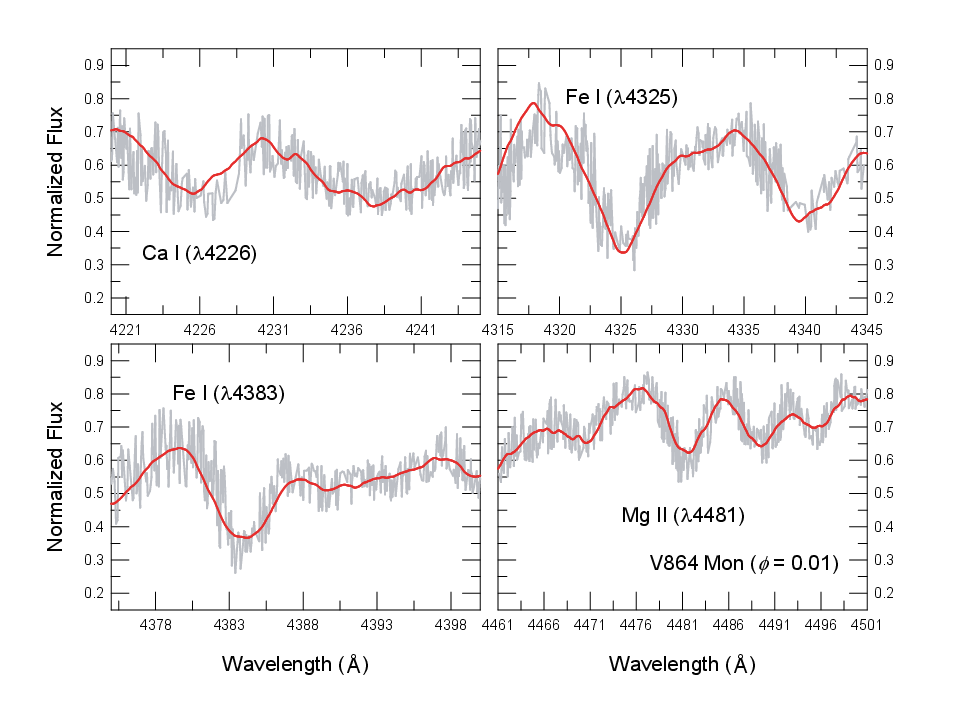}
\caption{Four spectral regions of the more massive secondary star, observed at an orbital phase of $\phi$ = 0.01.
The red line represents the synthetic spectrum with the best-fit parameters of $T_{\rm eff,2}$ = 5450 K and $v_2 \sin i$ = 192 km s$^{-1}$.}
\label{Fig2}
\end{figure}

\clearpage
\begin{figure}
\includegraphics[]{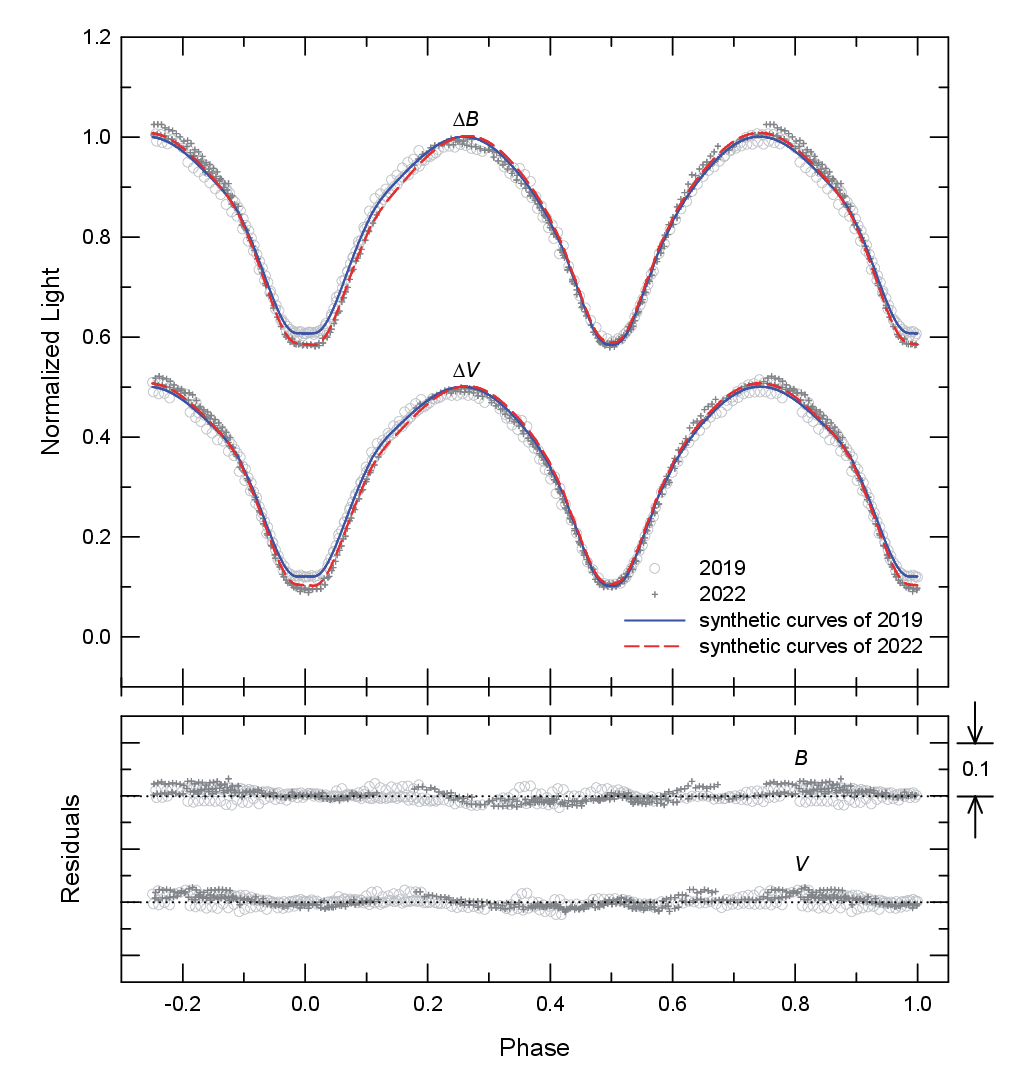}
\caption{Yearly $BV$ light curves of V864 Mon with fitted models.
The circles and plus symbols are the individual measures of the 2019 and 2022 seasons, respectively.
The solid and dashed lines represent the synthetic curves for each season obtained with the W-D run.
The lower panel shows the light residuals between measurements and theoretical models.}
\label{Fig3}
\end{figure}

\clearpage
\begin{figure}
\includegraphics[]{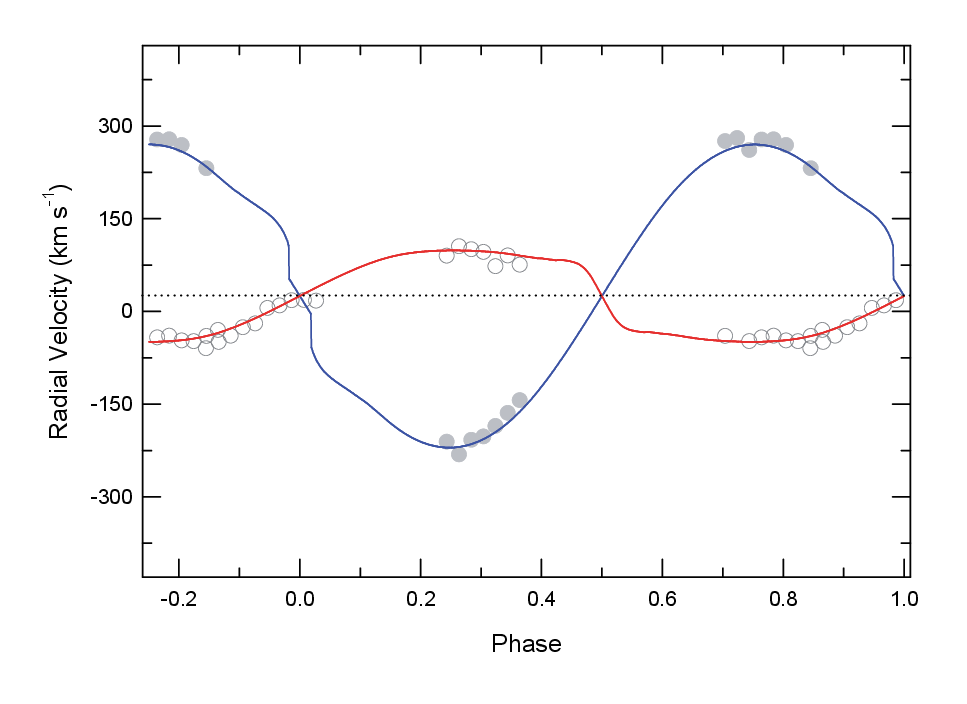}
\caption{RV curves of V864 Mon with fitted models.
The filled and open circles represent our double-lined RV measurements for the primary and secondary components, respectively.
The solid curves represent the results from a consistent light and RV curve analysis with the W-D code.
The dotted line represents the system velocity of $+$24.9 km s$^{-1}$.}
\label{Fig4}
\end{figure}

\clearpage
\begin{figure}
\includegraphics[width=1\columnwidth]{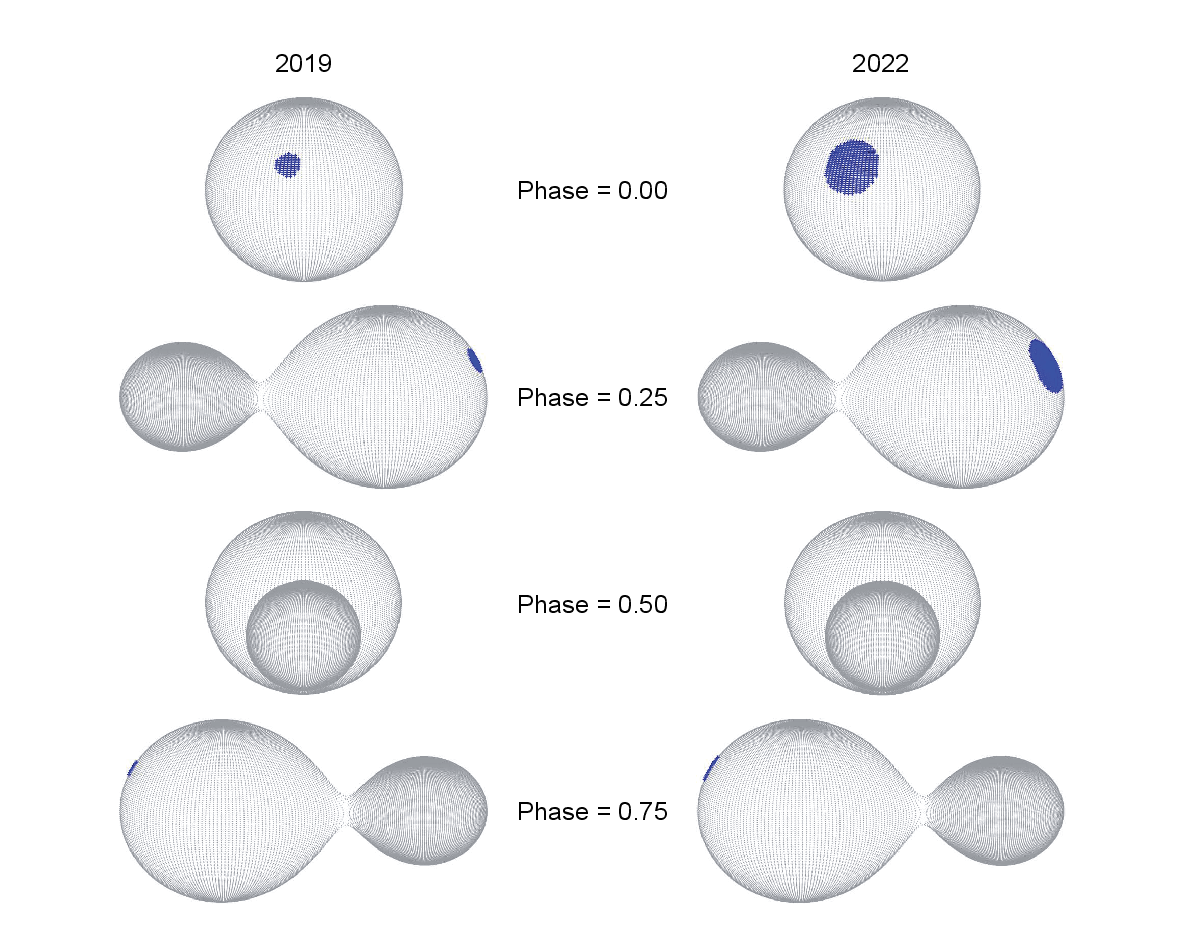}
\caption{2019 and 2022 3D models of V864 Mon with a cool spot on the secondary star at four different phases.}
\label{Fig5}
\end{figure}

\clearpage
\begin{figure}
\includegraphics[width=1\columnwidth]{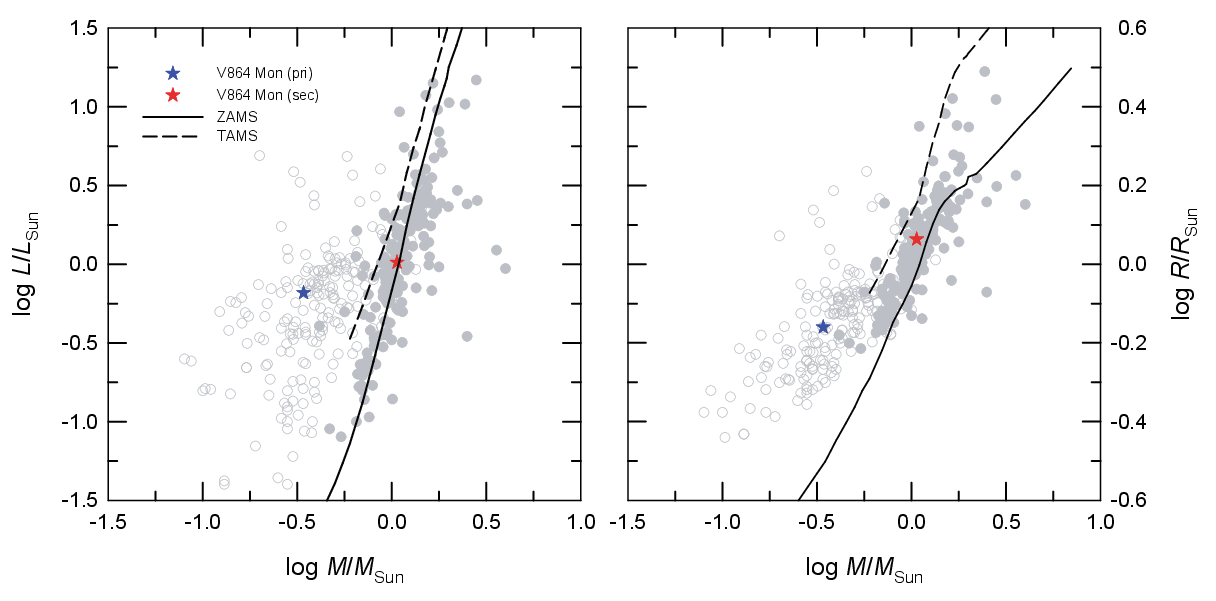}
\caption{Positions on the mass-luminosity ($M-L$) and mass-radius ($M-R$) diagrams for V864 Mon (star symbols) and other W-subtype of W UMa stars (circles; Latkovi\'c et al. 2021).
The filled and open circles represent the more massive and less massive components, respectively.}
\label{Fig6}
\end{figure}

\clearpage
\begin{table}
\tbl{Observed Visual and CCD timings of the minimum light for V864 Mon.}{%
\begin{tabular}{lcrrrccl}
\hline
HJD                   & Error                & Epoch                & $O$--$C_1$           & $O$--$C_2$           & Method               & Min                  & References                      \\ [-2.0ex]
(2,400,000+)          &                      &                      &                      &                      &                      &                      &                                 \\
\hline
52,667.755            &  \dots               &   $-$2354.0          &     0.00141          &  $-$0.00130          &  CCD                 &  I                   &  Wils \& Dvorak (2003)          \\
53,052.339            &  $\pm$0.003          &   $-$1281.0          &  $-$0.00517          &  $-$0.00640          &  VI                  &  I                   &  Vandenbroere et al. (2006)     \\
53,511.487            &  \dots               &         0.0          &  $-$0.00025          &  $-$0.00020          &  CCD                 &  I                   &  Kazarovets et al. (2006)       \\
54,888.7382           &  $\pm$0.0004         &      3842.5          &     0.00091          &     0.00250          &  CCD                 &  II                  &  Diethelm (2009)                \\
55,553.0770           &  \dots               &      5696.0          &  $-$0.00198          &  $-$0.00100          &  CCD                 &  I                   &  Nagai (2011)                   \\
55,621.7187           &  $\pm$0.0004         &      5887.5          &     0.00124          &     0.00210          &  CCD                 &  II                  &  Diethelm (2011)                \\
55,629.4237           &  $\pm$0.0009         &      5909.0          &     0.00009          &     0.00100          &  CCD                 &  I                   &  H\"ubscher et al. (2012)       \\
56,245.5576           &  $\pm$0.0007         &      7628.0          &     0.00053          &     0.00000          &  CCD                 &  I                   &  Ho{\v n}kov{\'a} et al. (2013) \\
56,308.9990           &  \dots               &      7805.0          &     0.00061          &  $-$0.00020          &  CCD                 &  I                   &  Nagai (2014)                   \\
56,354.3415           &  \dots               &      7931.5          &     0.00229          &     0.00140          &  CCD                 &  II                  &  Corfini et al. (2014)          \\
56,726.3868           &  $\pm$0.0012         &      8969.5          &     0.00190          &  $-$0.00030          &  CCD                 &  II                  &  H\"ubscher \& Lehmann (2015)   \\
56,728.3585           &  $\pm$0.0021         &      8975.0          &     0.00226          &     0.00000          &  CCD                 &  I                   &  H\"ubscher \& Lehmann (2015)   \\
56,994.1321           &  \dots               &      9716.5          &     0.00334          &     0.00000          &  CCD                 &  II                  &  Nagai (2015)                   \\
57,036.0662           &  \dots               &      9833.5          &     0.00166          &  $-$0.00190          &  CCD                 &  II                  &  Nagai (2016)                   \\
57,361.5194           &  $\pm$0.0001         &     10741.5          &     0.00449          &  $-$0.00070          &  CCD                 &  II                  &  Jury{\v s}ek et al. (2017)     \\
57,429.0830           &  \dots               &     10930.0          &     0.00488          &  $-$0.00060          &  CCD                 &  I                   &  Nagai (2017)                   \\
57,730.1632           &  \dots               &     11770.0          &     0.00765          &     0.00040          &  CCD                 &  I                   &  Nagai (2017)                   \\
57,776.038            &  \dots               &     11898.0          &     0.00398          &  $-$0.00350          &  CCD                 &  I                   &  Nagai (2018)                   \\
57,798.4425           &  $\pm$0.0012         &     11960.5          &     0.00689          &  $-$0.00070          &  CCD                 &  II                  &  Pagel (2018)                   \\
58,491.10556          &  $\pm$0.00021        &     13893.0          &     0.01264          &     0.00030          &  CCD                 &  I                   &  This Paper (2019)              \\
58,492.18086          &  $\pm$0.00024        &     13896.0          &     0.01267          &     0.00030          &  CCD                 &  I                   &  This Paper (2019)              \\
58,512.9732           &  \dots               &     13954.0          &     0.01633          &     0.00390          &  CCD                 &  I                   &  Nagai (2020)                   \\
58,539.3134           &  $\pm$0.0017         &     14027.5          &     0.01225          &  $-$0.00040          &  CCD                 &  II                  &  Pagel (2020)                   \\
58,539.4906           &  $\pm$0.0020         &     14028.0          &     0.01024          &  $-$0.00240          &  CCD                 &  I                   &  Pagel (2020)                   \\
58,578.02373          &  $\pm$0.00007        &     14135.5          &     0.01263          &  $-$0.00030          &  CCD                 &  II                  &  This Paper (2019)              \\
59,607.07203          &  $\pm$0.00023        &     17006.5          &     0.02128          &  $-$0.00060          &  CCD                 &  II                  &  This Paper (2022)              \\
59,608.14785          &  $\pm$0.00016        &     17009.5          &     0.02182          &  $-$0.00010          &  CCD                 &  II                  &  This Paper (2022)              \\
59,609.04513          &  $\pm$0.00011        &     17012.0          &     0.02304          &     0.00110          &  CCD                 &  I                   &  This Paper (2022)              \\
59,610.12053          &  $\pm$0.00011        &     17015.0          &     0.02316          &     0.00120          &  CCD                 &  I                   &  This Paper (2022)              \\
\hline
\end{tabular}}\label{tab:1}
\end{table}

\begin{table}
\tbl{Radial Velocities of V864 Mon.}{%
\begin{tabular}{lcrrrr}
\hline
HJD                    & Phase                  & $V_{1}$                & $\sigma_{1}$           & $V_{2}$                & $\sigma_{2}$           \\ [-2.0ex]
(2,450,000+)           &                        & km s$^{-1}$            & km s$^{-1}$            & km s$^{-1}$            & km s$^{-1}$            \\
\hline
8,500.0103             &  0.845                 &     231.6              &       9.1              &   $-$39.8              &       9.3              \\
8,500.0175             &  0.866                 &  \dots                 &  \dots                 &   $-$49.5              &      17.1              \\
8,500.0248             &  0.886                 &  \dots                 &  \dots                 &   $-$39.4              &       5.6              \\
8,500.0320             &  0.906                 &  \dots                 &  \dots                 &   $-$25.8              &       2.9              \\
8,500.0392             &  0.926                 &  \dots                 &  \dots                 &   $-$19.5              &       4.0              \\
8,500.0464             &  0.946                 &  \dots                 &  \dots                 &       5.5              &       3.5              \\
8,500.0537             &  0.966                 &  \dots                 &  \dots                 &       9.7              &       0.6              \\
8,500.0609             &  0.987                 &  \dots                 &  \dots                 &      18.0              &       5.7              \\
8,500.0681             &  0.007                 &  \dots                 &  \dots                 &      18.4              &       4.5              \\
8,500.0754             &  0.027                 &  \dots                 &  \dots                 &      17.2              &       2.4              \\
8,556.9492             &  0.704                 &     275.4              &       3.4              &   $-$39.6              &       3.7              \\
8,556.9564             &  0.724                 &     280.7              &       6.3              &  \dots                 &  \dots                 \\
8,556.9637             &  0.744                 &     261.2              &       3.5              &   $-$48.1              &       9.3              \\
8,556.9709             &  0.764                 &     278.1              &       5.3              &   $-$41.9              &      13.6              \\
8,556.9781             &  0.784                 &     278.3              &       2.5              &   $-$39.3              &       6.4              \\
8,556.9854             &  0.804                 &     269.1              &       5.4              &   $-$47.1              &       6.9              \\
8,556.9926             &  0.824                 &  \dots                 &  \dots                 &   $-$48.2              &       1.5              \\
8,556.9998             &  0.845                 &  \dots                 &  \dots                 &   $-$59.7              &       4.3              \\
8,557.0070             &  0.865                 &  \dots                 &  \dots                 &   $-$30.1              &       9.5              \\
8,558.9348             &  0.243                 &  $-$210.9              &       5.3              &      90.1              &      11.5              \\
8,558.9420             &  0.263                 &  $-$231.6              &       5.9              &     105.3              &       2.3              \\
8,558.9492             &  0.283                 &  $-$207.7              &       6.0              &     100.4              &       6.5              \\
8,558.9564             &  0.304                 &  $-$202.4              &       5.7              &      96.2              &       2.2              \\
8,558.9637             &  0.324                 &  $-$185.4              &       9.7              &      73.2              &       9.9              \\
8,558.9709             &  0.344                 &  $-$164.3              &      10.9              &      90.5              &       6.8              \\
8,558.9781             &  0.364                 &  $-$143.6              &       6.6              &      75.6              &       4.3              \\
\hline
\end{tabular}}\label{tab:2}
\end{table}

\begin{table}
\tbl{Binary Parameters of V864 Mon.}{%
\begin{tabular}{lcc}
\hline
Parameter                                & Primary             & Secondary                   \\
\hline
$i$ (deg)                                & \multicolumn{2}{c}{80.7 $\pm$ 0.3}                \\
$T$ (K)                                  & 5467  $\pm$ 94      & 5450  $\pm$ 94              \\
$\Omega$                                 & \multicolumn{2}{c}{6.696 $\pm$ 0.008}             \\
$\Omega_{\rm in}$                        & \multicolumn{2}{c}{6.777}                         \\
$\Omega_{\rm out}$                       & \multicolumn{2}{c}{6.156}                         \\
$r$ (pole)                               & 0.2715 $\pm$ 0.0006 & 0.4559 $\pm$ 0.0005         \\
$r$ (side)                               & 0.2835 $\pm$ 0.0007 & 0.4906 $\pm$ 0.0007         \\
$r$ (back)                               & 0.3205 $\pm$ 0.0011 & 0.5179 $\pm$ 0.0009         \\
$r$ (volume)$\rm ^a$                     & 0.2911 $\pm$ 0.0008 & 0.4874 $\pm$ 0.0007         \\
$f$ (fill-out)$\rm ^b$                   & \multicolumn{2}{c}{0.130}                         \\
\multicolumn{3}{l}{Spectroscopic:}                                                           \\
$a$ (R$_\odot$)                          & \multicolumn{2}{c}{2.37 $\pm$ 0.03}               \\
$\gamma$ (km s$^{-1}$)                   & \multicolumn{2}{c}{24.9 $\pm$ 2.1}                \\
$K_{1}$ (km s$^{-1}$)                    & \multicolumn{2}{c}{250.9 $\pm$ 3.9}               \\
$K_{2}$ (km s$^{-1}$)                    & \multicolumn{2}{c}{80.4 $\pm$ 2.6}                \\
$q$                                      & \multicolumn{2}{c}{3.122 $\pm$ 0.080}             \\
\hline
\multicolumn{3}{l}{Absolute Parameters:}                                                     \\
$M$ (M$_\odot$)                          & 0.34    $\pm$ 0.02  & 1.06    $\pm$ 0.04          \\
$R$ (R$_\odot$)                          & 0.69    $\pm$ 0.01  & 1.16    $\pm$ 0.02          \\
$\log$ $L/L_\odot$                       & $-$0.41 $\pm$ 0.03  & 0.03    $\pm$ 0.03          \\
$\log$ $g$ (cgs)                         & 4.29    $\pm$ 0.01  & 4.34    $\pm$ 0.01          \\
$M_{\rm bol}$ (mag)                      & $+$5.78 $\pm$ 0.08  & $+$4.68 $\pm$ 0.08          \\
BC (mag)                                 & $-$0.13             & $-$0.13                     \\
$M_{\rm V}$ (mag)                        & $+$5.91 $\pm$ 0.10  & $+$4.81 $\pm$ 0.10          \\
\hline
\end{tabular}}\label{tab:3}
\begin{tabnote}
\footnotemark[a]Mean volume radius calculated from the tables given by Mochnacki (1984). \\
\footnotemark[b]Fill-out factor ($f$) = ($\Omega_{\rm in}$--$\Omega$)/($\Omega_{\rm in}$--$\Omega_{\rm out}$),
where $\Omega_{\rm in}$ and $\Omega_{\rm out}$ represent the inner and outer critical Roche lobes, respectively.
\end{tabnote}
\end{table}

\begin{table}
\tbl{Luminosity and Spot Parameters for Each Dataset.}{%
\begin{tabular}{lcccc}
\hline
Years                                & \multicolumn{2}{c}{2019}                    & \multicolumn{2}{c}{2022}                    \\ [1.0mm] \cline{2-3} \cline{4-5}
                                     & Primary              & Secondary            & Primary              & Secondary            \\
\hline
$T_0$ (HJD)$\rm ^a$                  & \multicolumn{2}{c}{8,491.1050 $\pm$ 0.0001} & \multicolumn{2}{c}{8,491.1122 $\pm$ 0.0108} \\
$P$ (day)                            & \multicolumn{4}{c}{0.358426 $\pm$ 0.000001}                                               \\
$l_1$/($l_{1}$+$l_{2}$){$_{B}$}      & 0.2678 $\pm$ 0.0036  & 0.7322               & 0.2678 $\pm$ 0.0033  & 0.7322               \\
$l_1$/($l_{1}$+$l_{2}$){$_{V}$}      & 0.2666 $\pm$ 0.0028  & 0.7334               & 0.2666 $\pm$ 0.0026  & 0.7334               \\
Colatitude (deg)                     & \dots                & 67.00 $\pm$ 0.92     & \dots                & 68.03 $\pm$ 0.74     \\
Longitude (deg)                      & \dots                & 170.00 $\pm$ 0.93    & \dots                & 160.55 $\pm$ 1.34    \\
Radius (deg)                         & \dots                & 7.00 $\pm$ 0.44      & \dots                & 16.27 $\pm$ 0.40     \\
$T$$\rm _{spot}$/$T$$\rm _{local}$   & \dots                & 0.938 $\pm$ 0.021    & \dots                & 0.906 $\pm$ 0.004    \\
\hline
\end{tabular}}\label{tab:4}
\begin{tabnote}
\footnotemark[a]HJD 2,450,000 is suppressed.
\end{tabnote}
\end{table}


\end{document}